\documentclass[11pt,a4paper]{article}
\usepackage{jheppub}

\usepackage{slashed}
\usepackage{epsfig}
\usepackage{url}

\newcommand{\vect}[1]{\boldsymbol{#1}}
\def\vb{\begingroup\obeyspaces\u}
\def\u#1{\tt#1\endgroup}

\title{Double Parton Scattering Singularity in One-Loop Integrals} 
\author{Jonathan R. Gaunt,} 
\author{W. James Stirling} 
\affiliation{Cavendish Laboratory, University of Cambridge, \\J.J. Thomson Avenue, Cambridge
CB3 0HE, U.K.} 

\emailAdd{gaunt@hep.phy.cam.ac.uk} 
\emailAdd{wjs2@cam.ac.uk} 
\abstract{
We present a detailed study of the double parton scattering (DPS) singularity, which is a specific type of Landau 
singularity that can occur in certain one-loop graphs in theories with massless particles. A simple formula 
for the DPS singular part of a four-point diagram with arbitrary internal/external particles is derived in terms 
of the transverse momentum integral of a product of light cone wavefunctions with tree-level matrix elements.
This is used to reproduce and explain some results for DPS singularities in box integrals that have been obtained
using traditional loop integration techniques. The formula can be straightforwardly generalised to calculate the
DPS singularity in loops with an arbitrary number of external particles. We use the generalised version to explain 
why the specific MHV and NMHV six-photon amplitudes often studied by the NLO multileg community are not divergent
at the DPS singular point, and point out that whilst all NMHV amplitudes are always finite, certain MHV amplitudes
do contain a DPS divergence. It is shown that our framework for calculating DPS divergences in loop diagrams
is entirely consistent with the `two-parton GPD' framework of Diehl and Schafer for calculating proton-proton DPS cross 
sections, but is inconsistent with the `double PDF' framework of Snigirev.
}
\arxivnumber{1103.1888}
\dedicated{Cavendish--HEP--11/02}

\begin{document}
 
\maketitle

\section{Introduction} \label{sec:intro}

A necessary part of any one-loop calculation is the loop integration over the undetermined four-momentum $k$ in
each diagram contributing to the process considered. A loop integration will become singular if the $4$-dimensional 
real hypercontour over which the integration is performed becomes pinched by two (or more) poles associated with the
denominator factors in the integrand. Such singularities are known as Landau singularities, and they have been 
studied for some time \cite{AnalyticSMatrix}.

The denominator of a one-loop integral is equal to the product of propagator denominators in the associated
Feynman diagram, which is independent of the nature of the particles in the diagram. Thus, the Landau singularities
in a particular Feynman diagram are independent of the nature of the particles in it. The behaviour of the 
integral at a singular point can however be affected by the nature of the particles in the diagram, which determines
the numerator of the loop integral. If the numerator vanishes at the singular point, then the integral could be less
singular than expected there, or even finite.

\begin{figure}[b]
\centering
\includegraphics[scale=0.6]{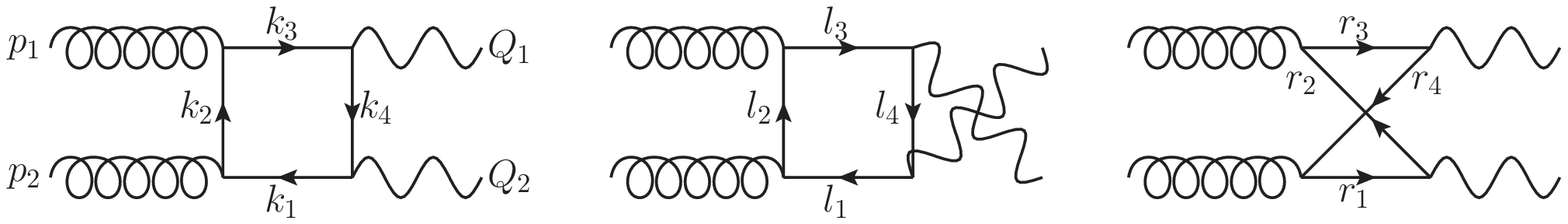}
\caption{\label{fig:ggZZboxes} Box topologies contributing to $gg \to ZZ$.}
\end{figure}

A relevant example of a one-loop calculation in which Landau singularities are encountered is $gg \to ZZ$ via
massless quark boxes. Three of the six box topologies contributing to this process are sketched in figure 
\ref{fig:ggZZboxes} -- the other three only differ by the direction of the arrow in the closed quark loop,
and give the same contributions as the boxes drawn.

Apart from mundane threshold singularities, the loops in figure \ref{fig:ggZZboxes} contain Landau singularities
that are associated with the initial state and loop particles being massless. In fact, all of the diagrams in
figure \ref{fig:ggZZboxes} contain at least one of these singularities for arbitrary values of the external
invariants. Every loop integral contains a collinear 
singularity (to be more precise, two collinear singularities), which is so named because it is associated
with a quark-antiquark pair attached to one of the external gluons becoming on-shell and collinear to that
gluon. The first and second loops in figure \ref{fig:ggZZboxes} also contain a soft singularity, which is associated 
with the four-momentum $k_2$ or $l_2$ shrinking to zero.

We shall refer to the final box topology in figure \ref{fig:ggZZboxes} as the crossed box due to its appearance 
when drawn with initial states on the left and final states on the right (although, for reasons of clarity, we will not draw it in this
way elsewhere in this paper -- see figure \ref{fig:genericbox} for example). This contains a singularity which is not shared
with the other two box topologies, and which only appears when the transverse momenta of the final states in the
centre of momentum frame, $\vect{Q_1}$ and $\vect{Q_2}$, are zero. This singularity is known as the double parton
scattering (DPS) singularity \cite{Nagy:2006xy}, and it is associated with all of the loop particles becoming on-shell
and collinear with the initial state gluons. The reason why the singularity is known as such is that it corresponds 
to the physical process in which two gluons each split to produce an on-shell, collinear quark-antiquark pair, and 
then the four resultant partons interact to produce two $Z$ bosons. The four partons interact in pairs from different 
gluons in two separate annihilation interactions, which is essentially the definition of a double parton scattering.

None of these singularities are restricted to the box diagrams. The conditions to have a collinear or soft singularity
in a one-loop diagram are well-documented \cite{Ellis:2007qk, Ninh:2008cz}. The double parton scattering singularity 
will occur for any one-loop diagram which satisfies the following criteria. First, the two initial state particles 
must be massless, and each of these initial state particles should be connected to two loop particles which are also 
massless. Then, the four massless loop particles should interact in two separate pairs, with particles from different 
initial state particles interacting. There is no restriction on the final state from each interaction, only that it 
should have total invariant mass squared which is timelike. Such diagrams will also generically contain collinear 
singularities.

The collinear and soft singularities in any loop diagram of the Standard Model are completely cancelled by the
numerator structure. This has to be the case -- otherwise the loop diagrams would be infinite for arbitrary
values of the external invariants, and one would get nonsensical infinite values for the differential cross
section of various processes. To what extent is the DPS singularity cancelled in Standard Model loops?

This answer to this question is of interest to two groups of people. The first of these is the NLO multileg 
community, who need to know where the singularities are in a loop integral, and how bad they are, to ensure 
(for example) accurate numerical evaluation of the loop integral \cite{Nagy:2006xy,Ossola:2007bb, Gong:2008ww}. It is also 
of interest to the multiple parton interaction (MPI) community, since the nature of the DPS singularity in one
loop diagrams determines whether part of these diagrams should be regarded as a leading order (LO) double
parton scattering process or not. If the infra-red DPS singularity in one-loop diagrams of the appropriate 
structure is not integrable at the cross section level, then the singularity should be absorbed into multiparton 
distributions, and a part of each loop diagram can be associated with LO DPS. On the other hand, if the DPS 
singularity is integrable at the cross section level, then the one-loop diagrams can simply be
regarded as contributing to single parton scattering processes.

The framework suggested by Snigirev for calculating the cross section for double parton scattering processes 
\cite{Snigirev:2003cq} anticipates that there should be an unintegrable DPS singularity in one-loop diagrams
of the appropriate structure. We show this explicitly in the following few paragraphs. Note that from 
henceforth, we shall refer to the framework of \cite{Snigirev:2003cq} as the `dPDF framework' (for reasons 
that will become clear shortly).

Very generally, one may write the cross section for the process $pp \to AB + X$ via double parton scattering 
as follows:
\begin{align} \label{DPSXsecgen}
\sigma^{D}_{(A,B)} \propto& \sum_{i,j,k,l}\int d^2\vect{r}\prod_{a=1}^{4}dx_a \Gamma_{ij}(x_1,x_2,\vect{r};Q_A^2,Q_B^2) \Gamma_{kl}(x_3,x_4,-\vect{r};Q_A^2,Q_B^2)
\\ \nonumber
\times& \hat{\sigma}_{ik \to A}(\hat{s} = x_1x_3s) \hat{\sigma}_{jl \to B}(\hat{s} = x_2x_4s)
\\ \nonumber
\propto& \sum_{i,j,k,l}\int d^2\vect{b}\prod_{a=1}^{4}dx_a \Gamma_{ij}(x_1,x_2,\vect{b};Q_A^2,Q_B^2) \Gamma_{kl}(x_3,x_4,\vect{b};Q_A^2,Q_B^2)
\\ \nonumber
\times& \hat{\sigma}_{ik \to A}(\hat{s} = x_1x_3s) \hat{\sigma}_{jl \to B}(\hat{s} = x_2x_4s)
\end{align}

The $\hat{\sigma}$ symbols represent parton-level cross sections. $\Gamma_{ij}(x_1,x_2,\vect{b};Q_A^2,Q_B^2)$
is the impact-parameter space two-parton GPD ($\vect{b}$-space 2pGPD), whilst $\Gamma_{ij}(x_1,x_2,\vect{r};Q_A^2,Q_B^2)$
is the transverse momentum space 2pGPD ($\vect{k}$-space 2pGPD). $\Gamma_{ij}(x_1,x_2,\vect{b};Q_A^2,Q_B^2)$ has a probability interpretation as 
the probability to find a pair of quarks in the proton with flavours $ij$, momentum fractions $x_1x_2$, and separated by
impact parameter $\vect{b}$, at scales $Q_A$ and $Q_B$ respectively \cite{Blok:2010ge, Diehl:2011tt}. The $\vect{k}$-space 2pGPD
is the Fourier transform of this with respect to $\vect{b}$, and has no probability interpretation.

Under the dPDF framework, it is assumed that $\Gamma_{ij}(x_1,x_2,\vect{b};Q_A^2,Q_B^2)$ may be approximately
factorised into the product of longitudinal and transverse pieces, with the longitudinal piece being given by
the double PDF (dPDF) object of \cite{Zinovev:1982be} for the case in which the two scales $Q_A$ and $Q_B$ are
equal. The transverse piece is typically taken to be flavour and scale independent:
\begin{equation} \label{2pGPDdecomp2dPDF}
\Gamma_{ij}(x_1,x_2,\vect{b};Q^2,Q^2) \xrightarrow[\text{framework}]{\text{dPDF}} D_p^{ij}(x_1,x_2;Q^2) F(\vect{b})
\end{equation}

According to \cite{Zinovev:1982be} the dPDFs evolve with $Q^2$ at LO according to:
\begin{align} \label{dbDGLAP}
Q^2\dfrac{dD^{{j_1}{j_2}}_p(x_1,x_2;Q)}{dQ^2} = \dfrac{\alpha_s(Q^2)}{2\pi} \Biggl[
\sum_{j'_1}\int_{x_1}^{1-x_2}\dfrac{dx'_1}{x'_1}D^{{j'_1}{j_2}}_p(x'_1,x_2;Q)P_{j'_1
\to j_1}\left(\dfrac{x_1}{x'_1}\right)
\nonumber\\
+\sum_{j'_2}\int_{x_2}^{1-x_1}\dfrac{dx'_2}{x'_2}D^{{j_1}{j'_2}}_p(x_1,x'_2;Q)P_{j'_2
\to j_2}\left(\dfrac{x_2}{x'_2}\right)
\nonumber\\
+\sum_{j'}D^{j'}_p(x_1+x_2;Q)\dfrac{1}{x_1+x_2}P_{j' \to j_1
j_2}\left(\dfrac{x_1}{x_1+x_2}\right) \Biggr]
\end{align}

The first two terms on the right hand side of \eqref{dbDGLAP} are associated with changes in the dPDF due to 
independent branching processes -- processes in which there are a pair of partons, one of which has the appropriate 
$x$ and flavour, and the other of which splits, either giving rise to the other parton of the appropriate $x$ and 
flavour, or removing it. The final inhomogeneous term represents the increase in the dPDF due to a single parton 
with momentum fraction $x_1+x_2$ splitting into a pair with the appropriate $x$ values and flavours. We call this 
the `sPDF feed term' for obvious reasons. The functions $P_{j\to j_1j_2}(x)$ that appear in this term are known as 
the $1\to 2$ splitting functions, and may be obtained trivially at LO from the real splitting parts of the usual 
splitting functions.

\begin{figure}
\centering
\includegraphics[scale=0.6]{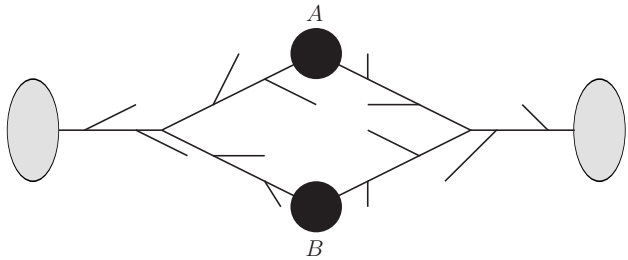}
\caption{\label{fig:dpsloops} A diagram that apparently contributes to the leading order DPS cross section according to the
‘dPDF framework’ (see text). The black circles are hard processes, the grey
blobs are protons, and the lines are partons.}
\end{figure}

Say we wish to calculate the cross section for a DPS process for which $Q_A=Q_B$ ($A=W^\pm,B=W^\pm$, for example).
It is clear that if we use \eqref{2pGPDdecomp2dPDF} in \eqref{DPSXsecgen} with the dPDFs at arbitrary scale
being obtained from some inputs at a low scale according to \eqref{dbDGLAP}, then the result for the cross
section will contain a term which contains the accumulated sPDF feed parts of two dPDFs being multiplied
together. Pictorially, this term corresponds to a sum of terms with the structure of figure \ref{fig:dpsloops}.

Since the dPDF framework includes figure \ref{fig:dpsloops} in the LO DPS cross section, this framework
predicts that the loop process of figure \ref{fig:dpsloops} should contain a piece which is 
proportional to $[\alpha_S \log(Q^2)]^N$ at the cross section level (where $N$ is the sum total of branchings that occur on either
side of the two hard processes $A$ and $B$). For such a piece to exist, each branching 
in the diagram has to be associated with a transverse momentum integration $\int d|\vect{k}|/|\vect{k}|$ 
at the cross section level -- even the vertices at which the branches on either side of the diagram split into two.
The leading log part of the cross section is then associated with the region in which the 
transverse momenta are strictly ordered along the branchings on either side of the diagram.

The $(\int d|\vect{k}|/|\vect{k}|)^2$ factor that is to be assigned to the two $1 \to 2$ branching vertices
in the diagram must originate in the DPS singularity of the loop integral, which must therefore be infra-red 
divergent like a logarithm squared at the cross section level prior to subtractions. At the amplitude level, the DPS singularity
in the loop integral would have to be $(\log(p_T^2)/p_T^2)^{0.5}$ to give such a divergence at the cross section level
(where $p_T$ is equal to the absolute value of the transverse momentum of all of the particles on the upper (or the lower) 
leg of the loop, in the centre of momentum frame of the two particles initiating the loop). 

On the other hand, explicit calculations of certain four- and six-point amplitudes within the Standard Model, which
contain diagrams of the appropriate character, show that these diagrams at least do not contain unintegrable DPS
singularities \cite{Glover:1987nx, Glover:1988rg, Binoth:2006mf, Bernicot:2008nd, Bern:2008ef}. The general impression 
one gets from the literature is that the DPS singularity in any Standard Model one-loop diagram is completely
cancelled, such that the loop integral is finite in the limit $p_T \to 0$. 

In this paper, we present a detailed and general study of the DPS singularity in one-loop integrals. To begin with,
we only focus on the four-point diagram that can contain a DPS singularity -- i.e. the crossed box. In section 
\ref{sec:BoxSingularities}, we present results for the DPS singular parts of certain crossed box diagrams, which
have been obtained using traditional loop integral techniques. Some of these have been extracted from the available
literature, whilst others are derived by us. We show that in certain Standard Model crossed box diagrams, the 
DPS singularity is not completely cancelled, but is instead relegated to an integrable logarithm.

It is not efficient to evaluate the DPS singular part of a given four-point (or $n$-point) diagram using 
traditional loop integral techniques, since such techniques involve the full calculation of the rest of the
integral (which we are not interested in here). Further, we can gain little insight from these techniques
as to why a particular box integral has a DPS divergence of a given nature (in particular, why the Standard
Model boxes have a DPS divergence which is at most a logarithm of $\vect{Q_2}$). In section 
\ref{sec:PhysicalInvestigation} we derive a framework for evaluating the DPS singular part of a crossed box 
diagram which only requires the calculation of simple leading order light-cone wavefunctions and tree-level matrix elements. 
Using this framework, we reproduce and provide physical explanations for
 all of the crossed box results found in section \ref{sec:BoxSingularities}.

The framework that we derive for calculating the DPS divergence of a crossed box has the advantage that it
is very easily generalised to loops with arbitrary numbers of external particles. In section \ref{sec:MoreLegs},
we use the generalised framework to check and generalise the results of \cite{Bern:2008ef,Bernicot:2008nd} for 
the DPS divergence in six-photon helicity amplitudes. We also use it to determine the nature of the DPS singularity
in figure \ref{fig:dpsloops}, and compare the result with the predictions of the dPDF framework. Based on the 
outcome of this comparison, we comment on the validity of the dPDF framework versus the `two-parton GPD' framework
for calculating proton-proton DPS recently introduced by Diehl and Schafer \cite{Diehl:2011tt}.

\section{Singularities in the Crossed Box} \label{sec:BoxSingularities}

\begin{figure}
\centering
\includegraphics[scale=0.6]{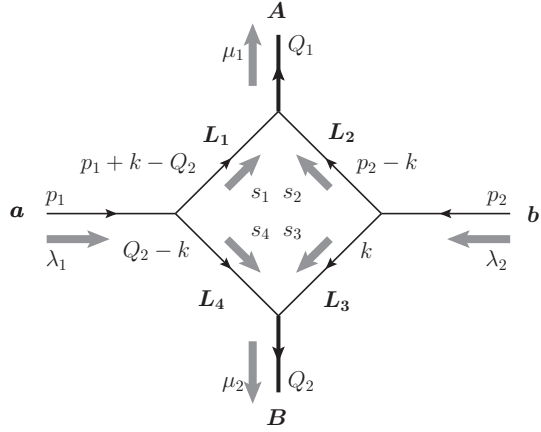}
\caption{\label{fig:genericbox} The crossed box topology, with annotations that demonstrate our labelling conventions
for the particle names, helicity and momenta. The particle names are written in bold in this figure, whilst the helicity
labels are accompanied by grey arrows. The arrows on the lines merely indicate 
the direction of momentum flow, and do not necessarily signify a fermion line. The thin lines represent massless particles, 
whereas the thick lines represent particles with invariant mass squared equal to $M^2$.}
\end{figure}

We consider a generic crossed box diagram with the momenta and helicities labelled as in figure \ref{fig:genericbox} (note
that any or all of the helicities could be zero in general). For the moment, we
do not specify the nature of the external and loop particles. We do however impose the conditions that are necessary 
for the crossed box to contain a DPS singularity -- namely, that
the incoming particles (with momenta $p_1$ and $p_2$) should be on-shell and massless, whilst the outgoing particles
(with momenta $Q_2$ and $Q_2$) should either be on-shell and massive, or off-shell such that $Q_1^2$ and $Q_2^2>0$.
For the purposes of calculational simplicity, the squared four momenta of $Q_1$ and $Q_2$ shall be taken to be equal
in all of the boxes studied. The common four momentum squared $Q_1^2=Q_2^2$ will be denoted by $M^2$. Further, we work 
at all times in the centre of momentum frame, and choose the $z$ axis to be aligned with the spatial part of $p_1$. We define:
\begin{align}
s \equiv (p_1+p_2)^2 \quad
t \equiv (p_1-Q_1)^2 \quad
u \equiv (p_1-Q_2)^2
\end{align}
The $d$ dimensional loop integral associated with the crossed box has the following generic form:
\begin{align} \label{genericbox} 
L =& \int d^{d}k \dfrac{\mathcal{N}}{[k^2+i\epsilon][(k-Q_2)^2+i\epsilon][(p_1+k-Q_2)^2+i\epsilon][(p_2-k)^2+i\epsilon]}
\end{align}

\begin{figure}
\centering
\includegraphics[scale=0.59]{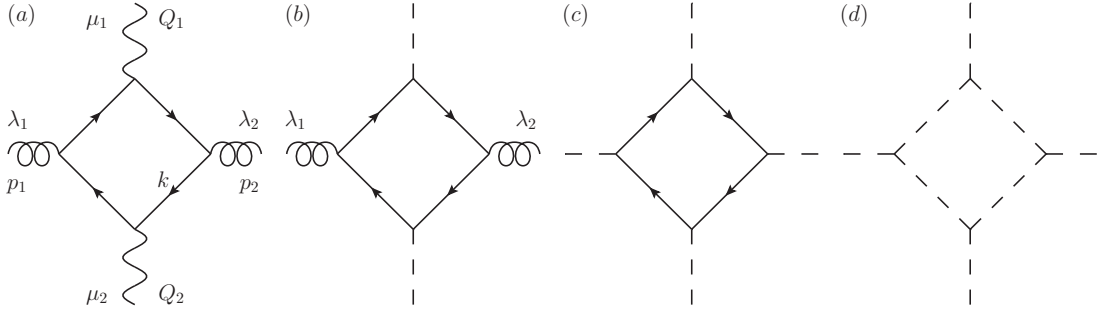}
\caption{\label{fig:boxes} The crossed boxes that we shall consider as examples in this paper. The helical lines
represent gluons, the wavy lines Z bosons, the dashed lines scalars, and the lines with arrows represent fermions.}
\end{figure}

The nature of the external and loop particles determines the numerator factor $\mathcal{N}$, but not the 
denominator. $L$ is defined such that $\mathcal{N}$ only includes the trace structure
of the crossed box amplitude, and does not include overall factors such as coupling constants and colour
factors. For future reference, we write here the numerator factors for each of the specific crossed boxes 
that we will consider as examples in this paper, and which are drawn in figure \ref{fig:boxes}:
\begin{align}
\mathcal{N} = \begin{cases}
              \mathrm{Tr}[\slashed{\varepsilon}^*_{\mu_2}\slashed{k}\slashed{\varepsilon}_{\lambda_1}(\slashed{p}_2-\slashed{k})
              \slashed{\varepsilon}^*_{\mu_1}(\slashed{p}_1 +\slashed{k}-\slashed{Q}_2)
              \slashed{\varepsilon}_{\lambda_2}(\slashed{Q}_2-\slashed{k})] & \text{for Fig \ref{fig:boxes}(a)} \\
              \mathrm{Tr}[\slashed{k}\slashed{\varepsilon}_{\lambda_1}(\slashed{p}_2-\slashed{k})
              (\slashed{p}_1 +\slashed{k}-\slashed{Q}_2)
              \slashed{\varepsilon}_{\lambda_2}(\slashed{Q}_2-\slashed{k})] & \text{for Fig \ref{fig:boxes}(b)} \\
              \mathrm{Tr}[\slashed{k}(\slashed{p}_2-\slashed{k})
              (\slashed{p}_1 +\slashed{k}-\slashed{Q}_2)
              (\slashed{Q}_2-\slashed{k})] & \text{for Fig \ref{fig:boxes}(c)} \\
              1 & \text{for Fig \ref{fig:boxes}(d)}
              \end{cases}
\end{align}
where $\epsilon_\lambda$ is the polarisation vector corresponding to helicity $\lambda$ ($\lambda = \pm 1$ for gluons, 
and $\pm 1, 0$ for Z bosons). The numerator factor for figure \ref{fig:boxes}(a) written above is of course not the 
initial expression you would write down, which would contain contain factors of $v_q + a_q \gamma^5$ before
$\slashed{\varepsilon}^*_{\mu_1}$ and $\slashed{\varepsilon}^*_{\mu_2}$ (where we use the notation of \cite{Glover:1988rg} 
-- $v_q$ ($a_q$) is the vector (axial) coupling of the quarks in the loop to $Z$ bosons). However, the terms in this 
initial expression proportional to $v_qa_q$ cannot contribute to the loop integral according to charge conjugation
invariance \cite{Glover:1988rg}, whilst the terms proportional to $v_q^2$ and $a_q^2$ can both be shown to have the 
 trace structure written above (some anticommutation of the $\gamma^5$ matrices plus use of $(\gamma^5)^2=1$ is 
required in the latter case). Thus, the numerator of the $gg \to ZZ$ loop integral is equal to the above trace structure,
times some overall coupling constant which we drop.

We recall that a crossed box contains collinear and double parton scattering Landau singularities.
In the case of $\phi^3$ theory in four dimensions, these nominal singularities both correspond to actual
infinite values of the integral. We may calculate the most divergent part of the crossed box in this theory
(drawn in figure \ref{fig:boxes}(d)) using the elegant method presented in \cite{Ninh:2008cz}. Note that we must work 
in  $4+2\varepsilon$ dimensions to regulate the collinear divergence which appears for arbitrary values of the kinematic 
invariants. The result of the calculation is:
\begin{align} \nonumber
L_{\phi,4D} &= -\dfrac{4\pi^3}{(ut-M^4)\varepsilon} + \text{less divergent terms}
\\  \label{Ninh4D}
            &= -\dfrac{4\pi^3}{s\vect{Q_2}^2\varepsilon} + \text{less divergent terms}
\end{align}
where $\vect{Q_2}^2$ is the transverse momentum squared of the second massive particle (= $ut-M^4$ = $\vect{Q_1}^2$).
We note that the result obtained here corresponds to minus the expression presented in equation (4.101) of 
\cite{Ninh:2008cz} -- the latter result is out by a factor of (-1) because the two poles in equation (4.99)
of \cite{Ninh:2008cz} are on the wrong sides of the real axis. One observes the appearance of a factor
$1/\varepsilon$ in the expression \eqref{Ninh4D} which corresponds to the collinear singularity and is 
infinite in the limit $\varepsilon \to 0$, and a factor $1/\vect{Q_2}^2$ which corresponds to the DPS
singularity and is infinite in the limit $\vect{Q_2} \to 0$. A critical point to note is that the DPS
singular part of the 4D scalar crossed box is not integrable -- that is, if ones takes its modulus
squared and integrates it over the final state phase space, then one obtains an infinite contribution
to the cross section (the result is proportional to $d\vect{Q_2}^2/\vect{Q_2}^4$).

In more complex four dimensional theories, there exists the possibility that the collinear and DPS
singularities in crossed box integrals may exhibit less singular behaviour, due to the fact that there
is now a nontrivial numerator factor $\mathcal{N}$ which may vanish at the singular points. Indeed,
this seems to be the case for every Standard Model crossed box one can construct obeying
the appropriate conditions (i.e. $p_1^2=p_2^2=0$, massless particles in the loop, and $Q_1^2$ and $Q_2^2 > 0$). 
In all of these boxes, the collinear divergence vanishes, and the DPS singularity is relegated to a 
logarithm in $\vect{Q_2}^2$ at most.

Let us give some examples of this logarithmic behaviour of the DPS singularity drawing from the
established literature. The first example we shall consider is $gg \to HH$ via a massless quark loop.
Glover and van der Bij have calculated the crossed box integral for this process \cite{Glover:1987nx}.
However, they only present results for the helicity matrix elements calculated using a general 
quark mass $m_q$ in the loop. The helicity matrix elements are equal to the sum over loop integrals 
for the six different loop topologies contributing to $gg \to HH$, all multiplied by various factors 
(coupling constants, colour factors). We can nevertheless extract the leading low $\vect{Q_2}$
behaviour of a single $gg \to HH$ crossed box from these results as follows. First, we strip the 
multiplying factors from the helicity matrix elements to obtain expressions for the loop integrals summed 
over topologies. This turns out to require extreme care since the authors of \cite{Glover:1987nx} have chosen
to factor some constants out of their matrix elements and into their expression for $d\sigma/dt$. Then,
we take the expansions for the scalar loop integrals in the low $m_q$ limit (found in Appendix B of
\cite{Glover:1988rg}), insert them into these expressions, and take $m_q \to 0$ to obtain the sums 
over loop topologies for the massless quark case\footnote{An alternative approach would be to insert 
the dimensionally regulated scalar loop integrals with massless internal lines found in \cite{Ellis:2007qk}. 
One has to exercise some care in analytically continuing the results of \cite{Ellis:2007qk} to the region of present
interest however -- see the detailed discussion in \cite{Duplancic:2000sk}. Of course, either method gives the same result.}.
Such a limit is perfectly well defined since the loop integrals do not contain collinear singularities. Finally, we isolate
any low $\vect{Q_2}$  divergences in the resulting expressions -- these can be equated to twice 
the leading low $\vect{Q_2}$ behaviour of the relevant crossed boxes. The reason for this is that
only the crossed box loop integral can contain a DPS singularity, and there are two crossed box
topologies that contribute equally to $gg \to HH$.

Performing this procedure, we find that there are two helicity configurations for which the crossed 
box diverges as $\vect{Q_2} \to 0$ -- these are the $++$ and $--$ configurations.
 The corresponding leading low $\vect{Q_2}$ behaviour of the crossed box integral with either of 
these helicity configurations is:
\begin{align} \label{GloverHiggs}
L_{DPS}(++) = L_{DPS}(--) = -\dfrac{8 M_H^2 \pi^3 \log(\vect{Q_2}^2)}{s}
\end{align}

This result may be directly checked by decomposing the  $gg \to HH$ crossed box loop 
integral to scalar integrals using \vb{FeynCalc} \cite{Mertig1991345}, before inserting the low $m_q$ 
expansions of the scalar integrals and taking $m_q \to 0$. We obtain the same expression using this
method.

Another example of a crossed box satisfying the appropriate conditions is $gg \to ZZ$, again via
a massless quark loop. Glover and van der Bij have calculated the loop integrals for this process
as well \cite{Glover:1988rg}. As in the $gg \to HH$ case, they only present amplitudes
summed over all box topologies and for the case of general quark mass -- however we can
extract the leading low $\vect{Q_2}$ behaviour of the $m_q=0$ $gg \to ZZ$ crossed boxes 
from these results using the same technique as was applied in the $gg \to HH$ case.

We denote the helicity configuration in a crossed box integral by $\lambda_1\lambda_2\mu_1\mu_2$ where $\lambda_1$
and $\lambda_2$ correspond to the helicities of gluons 1 and 2, and $\mu_1$ and $\mu_2$ correspond to the
helicities of Z bosons 1 and 2. Then only the $++++$,$----$, $++--$ and $--++$ integrals are
divergent in the limit $\vect{Q_2} \to 0$:
\begin{align} \label{GloverZboson}
L_{DPS}(++++) &= L_{DPS} = (----) = \dfrac{8\pi^3\left[s-2M_Z^2+s\sqrt{1-4M_Z^2/s}\right]\log(\vect{Q_2}^2)}{s}
\\
L_{DPS}(++--) &= L_{DPS} = (--++) = \dfrac{8\pi^3\left[s-2M_Z^2-s\sqrt{1-4M_Z^2/s}\right]\log(\vect{Q_2}^2)}{s}
\end{align}

Unfortunately we cannot check this result using \vb{FeynCalc} as it requires the Passarino-Veltman reduction 
\cite{Passarino:1978jh} of tensor integrals of index 4, which \vb{FeynCalc} cannot handle. We remark in
passing that the same results for $L_{DPS}$ are obtained if the final state $Z$ bosons are replaced by
off-shell photons (with $Q_1^2 = Q_1^2 = M^2$) or $W$ bosons, except that $M_Z$ in \eqref{GloverZboson} should be
replaced by $M$ or $M_W$. The coupling constant factor that multiplies $L_{DPS}$ in the full expression
for the amplitude is $(v_q^2+a_q^2)g_Z^2g_s^2$ in the $gg \to ZZ$ case. The coupling constant factor 
for $gg \to \gamma^*\gamma^*$ may be obtained from this by setting $v_q=Q_q,a_q=0$ and replacing $g_Z$ 
by $e$, whilst that for $gg \to W^+W^-$ is obtained by setting $v_q = -a_q = 1$ and replacing $g_Z$ by $g_w/(2\sqrt{2})$.

Despite assertions to the contrary that exist in the literature \cite{Binoth:2006mf},
some of the Standard Model crossed box loop integrals are divergent in the limit $\vect{Q_2} \to 0$. 
However, a key point is that they are not sufficiently divergent to cause the cross section for 
$gg \to ZZ$ or $gg \to HH$ to diverge ($\int d\vect{Q_2}^2 \log^2(\vect{Q_2}^2)$ = finite). In
particular, the log squared singularity in the cross section that would be anticipated by 
Snigirev \cite{Snigirev:2003cq} does not exist.

The phenomenon by which the numerator of Standard Model crossed box integrals always vanishes at the DPS
singular point such that the DPS singularity is integrable demands explanation. In \cite{Ninh:2008cz,Bern:2008ef}, it
is suggested that `gauge dynamics' is the physical cause. If this is the case, then we should not see
similar behaviour in a 4 dimensional theory which does not have a gauge symmetry -- for example, scalar
gluon theory (also known as massless Yukawa theory). Therefore, to test the hypothesis of \cite{Ninh:2008cz}, 
we examined the crossed box loop integral associated with the process $g_sg_s \to g_s^* g_s^*$ in 
scalar gluon theory, where the final state scalars are off-shell by the same timelike amount. The 
Feynman diagram corresponding to the integral is figure \ref{fig:boxes}(c).

To calculate the leading low $\vect{Q_2}$ behaviour of the $g_sg_s \to g_s^* g_s^*$ crossed box loop 
integral, we use two methods. First, we perform the Passarino-Veltman reduction of the integral `by hand' 
in \vb{Maple} \cite{Maple13}, before inserting the expansions of the scalar integrals for small loop particle
mass found in \cite{Glover:1988rg}, and then taking the limit of zero loop particle mass. The other approach
involves using \vb{FeynCalc} to perform the Passarino-Veltman reduction. Both approaches return the same
result:
\begin{align} \label{GauntYukawa}
L_{DPS} = 4\pi^3\log(\vect{Q_2}^2)
\end{align}

We see that the singular behaviour of this box is exactly the same as the Standard Model boxes -- i.e.
the DPS singularity becomes an integrable logarithm (and the collinear singularity disappears). This 
example indicates that we cannot uniquely associate a logarithmic DPS singularity with gauge theories,
and therefore that gauge symmetries are not the direct driving force behind this behaviour.

Although the scalar crossed box integral in four dimensions has a very different singular behaviour
to the Standard Model and Yukawa boxes, the same integral in {\em six} dimensions (corresponding to
6D $\phi^3$ theory) has exactly the same singular behaviour as the 4D SM and Yukawa boxes. We can 
calculate the most singular part of the 6D scalar box by applying the method found in section
4.6.2 of \cite{Ninh:2008cz} to $D=6$. In this case, we do not need to deform the number of dimensions
to $D=6+2\varepsilon$ since there are no collinear singularities in the integral.

Repeating the steps (4.95)-(4.97) of \cite{Ninh:2008cz} with $D=6$, we obtain:
\begin{equation}
L_{DPS,\phi,6D} = \pi^3 i \int_0^1 \dfrac{d\alpha d\beta}{[s\alpha+u-M^2-i\epsilon]\beta+[(u-M^2)\alpha-u-i\epsilon]}
\end{equation}
where $\alpha$ and $\beta$ are Feynman parameters. It is simple to perform the integration over $\beta$, which
gives:
\begin{align}
L_{DPS,\phi,6D} = \pi^3 i \int_0^1 \dfrac{d\alpha}{s\alpha+u-M^2-i\epsilon}&\bigl[\ln(s\alpha-M^2+(u-M^2)\alpha-i\epsilon)
\\ \nonumber
&-\ln((u-M^2)\alpha - u -i\epsilon)\bigr]
\end{align}

The real part of this integral is finite as $\vect{Q_2} \to 0$, and so for the purposes of extracting the
leading low $\vect{Q_2}$ singularity we can ignore it:
\begin{align} \nonumber
L_{DPS,\phi,6D} \simeq& \pi^3 i \int_0^1 \dfrac{d\alpha}{s\alpha+u-M^2}\left[-i\pi\Theta(M^2-s\alpha-(u-M^2)\alpha)
+i\pi\Theta(u-(u-M^2)\alpha)\right]
\\ \label{Ninh6D}
=& \dfrac{2\pi^4\log(\vect{Q_2}^2)}{s}
\end{align}
As asserted, the 6D scalar box has a logarithmic DPS singularity in its crossed box.

There must exist some characteristic that is common to 6D scalar theory, 4D scalar gluon theory, and
the Standard Model that ensures that the leading DPS singularity in these theories is converted from
a single inverse power of $\vect{Q_2}^2$ to a logarithm (and that the collinear singularity vanishes).
Using traditional techniques for handling loops, it is exceedingly difficult to elucidate the mechanism
by which this occurs. The reason for this is that we lose contact with the original structure of the 
loop integral when we start introducing Feynman parameters (and, in the Yukawa and SM cases, even before
this when we perform the Passarino-Veltman reduction). In the next section, we shall introduce a 
technique for directly calculating the portion of a crossed box loop integral which contains the DPS
singular point when $\vect{Q_2}=0$ (i.e. the point at which all of the internal lines go on shell). 
As the evaluation of the portion of the integral is direct, neither Passarino-Veltman reduction nor 
introduction of Feynman parameters needs to be performed. By use of this method, we will discover 
the physical origin of the logarithmic DPS singularity in 6D scalar theory, 4D scalar gluon theory, and
the Standard Model.

\section{Physical Investigation of the Crossed Box} \label{sec:PhysicalInvestigation}

We would like to investigate the nature and origin of the part of the amplitude $L$ which is most singular 
as the transverse momenta of the produced particles goes to zero. This part of the amplitude is associated 
with the region of the loop integration in which the transverse part of the loop variable, 
$\vect{k}$, is small (i.e. much less than $ \sqrt{s}$ and masses of produced particles). 
The reason for this is that, when the transverse momentum of the produced particles 
is zero, the small $\vect{k}$ region contains the point in which all four of the loop particles go on shell
simultaneously.

Therefore we study the contribution to $L$ coming from the small $\vect{k}$ region in the case in which the 
transverse momenta of the produced particles is also small. The method we use is similar to that described in
section V of \cite{Liu:1997kia}, although we fix some errors and address some subtleties of which the author
of \cite{Liu:1997kia} did not seem to be aware.

To begin, we apply a Sudakov decomposition to all of the vectors in \eqref{genericbox}. We define lightlike
vectors $n$ and $p$ as follows:
\begin{align}
p = \dfrac{1}{\sqrt{2}}(1,0,0,1) \qquad n = \dfrac{1}{\sqrt{2}}(1,0,0,-1)
\end{align}

An arbitary four vector $A$ may be written in terms of these vectors plus a transverse part $\vect{A}$ (which
only has $x$ and $y$ components) as follows:
\begin{equation}
A = A^+p+A^-n+\vect{A} 
\end{equation}
such that:
\begin{equation}
 A\cdot B = A^+B^- + A^-B^+ - \vect{A}\cdot\vect{B}
\end{equation}

Writing out all of the four momenta in \eqref{genericbox} in terms of $n$, $p$, and a transverse part, 
\eqref{genericbox} becomes:
\begin{align} \label{LCbox}\nonumber
L =& \int d^{d-2}\mathbf{k}dk^+dk^- \dfrac{\mathcal{N}}{(2k^+k^--\mathbf{k}^2+i\epsilon)[2(k^+-Q_2^+)(k^--Q_2^-)-(\mathbf{k-Q_2})^2+i\epsilon]}
\\
\times& \dfrac{1}{[2(k^++Q_1^+)(k^--Q_2^-)-(\mathbf{k-Q_2})^2+i\epsilon][2k^+(k^--Q_1^--Q_2^-)-\mathbf{k}^2+i\epsilon]}
\end{align}

In deriving \eqref{LCbox}, we have used the fact that, in our chosen reference frame for which $p_1 \propto p$,
$p_2 \propto n$, conservation of four momentum implies:
\begin{equation}
p_1 = (Q_1^++Q_2^+)p \qquad p_2 = (Q_1^-+Q_2^-)n
\end{equation}

In the following discussion, an important point to bear in mind is that $Q_i^+$ and $Q_i^-$ are always
positive (provided the masses of the produced particles are not zero).

The part of $L$ that we are interested in is the low $\vect{k}$ portion of the integral, which we shall
denote as $L_{DPS}$. Our strategy to evaluate $L_{DPS}$ will be to perform the $k^-$, $k^+$ and 
$\vect{k}$ integrals in that order, making copious use of the fact that $\vect{k}$ in the 
integration is small.

The $k^-$ integration is straightforward. When $0<k^+<Q_2^+$, only the second $k^-$ pole in the 
denominator lies on the upper half complex plane, so we close the contour on the upper half
plane and pick up the pole at:
\begin{equation}
k_2^- = Q_2^- + \dfrac{(\vect{k-Q})^2}{2(k^+-Q_2^+)}
\end{equation}

When $-Q_1^+<k^+<0$, only the third $k^-$ pole in the denominator is located in the lower half 
complex plane, so in this case we close the contour on the lower half plane and pick up the 
pole at:
\begin{equation}
k_3^- = Q_2^- + \dfrac{(\vect{k-Q})^2}{2(k^++Q_1^+)}
\end{equation}

Finally, when $k^+ < -Q_1^+$ or $k^+ > Q_2^+$, all of the poles lie on one side of the real axis,
so we close the contour on the other side and get zero for the value of the integral. Putting
it all together, we find that the result of the $k^-$ integration is the following:
\begin{align} \label{intk-}
L_{DPS} =& -2\pi i\int_{|\vect{k}|\ll Q_i^+,Q_i^-} d^{d-2}\vect{k}\int_{-Q_1^+}^{0}dk^+ \dfrac{\mathcal{N}\mid_{k^-=k^-_3}}
{\left(2k^+Q_2^- + \tfrac{k^+(\vect{k-Q_2})^2}{(k^++Q_1^+)}-\vect{k}^2+i\epsilon\right)} \\ \nonumber
\times& \dfrac{1}{2(-Q_1^+-Q_2^+)(\vect{k-Q_2})^2
\left(-2k^+Q_1^- + \tfrac{k^+(\vect{k-Q_2})^2}{(k^++Q_1^+)}-\vect{k}^2+i\epsilon\right)} \\ \nonumber
+& 2\pi i\int_{|\vect{k}|\ll Q_i^+,Q_i^-} d^2\vect{k}\int_{0}^{Q_2^+}dk^+ \dfrac{\mathcal{N}\mid_{k^-=k^-_2}}
{\left(2k^+Q_2^- + \tfrac{k^+(\vect{k-Q_2})^2}{(k^+-Q_2^+)}-\vect{k}^2+i\epsilon\right)} \\ \nonumber
\times& \dfrac{1}{2(Q_1^++Q_2^+)(\vect{k-Q_2})^2
\left(-2k^+Q_1^- + \tfrac{k^+(\vect{k-Q_2})^2}{(k^+-Q_2
^+)}-\vect{k}^2+i\epsilon\right)}
\end{align}

We note that the terms $k^+(\vect{k-Q_2})^2/(k^++Q_1^+)$ and $k^+(\vect{k-Q_2})^2/(k^+-Q_2^+)$ appear
in some of the denominator factors. These terms are negligible except where $k^+ \sim -Q_1^+$ or
$k^+ = Q_2^+$. However, the region of $k^+$ which is relevant to the leading $\vect{Q_2}$ singularity
in $L$ is $|k^+| \ll Q_i^+,Q_i^-$. This is because, when $\vect{Q_2}$ vanishes, the configuration
in which all of the loop particles are on shell corresponds to $k = Q_2^-n$ (i.e. $k^+ = 0$).
Therefore, for the purposes of finding the leading singularity in $L$, we can drop the 
$k^+(\vect{k-Q_2})^2/(k^++Q_1^+)$ and $k^+(\vect{k-Q_2})^2/(k^+-Q_2^+)$ terms in the denominator.
For similar reasons, we can replace $k_2^-$ and $k_3^-$ by $Q_2^-$ and set $k^+ = 0$ in the
numerator. Then, $L_{DPS}$ becomes:
\begin{align} \label{intk-2}
L_{DPS} \simeq& \dfrac{2\pi i}{2(Q_1^++Q_2^+)}\int_{|\vect{k}|\ll Q_i^+,Q_i^-} 
\dfrac{d^{d-2}\vect{k}}{(\vect{k-Q_2})^2}  \\ \nonumber
\times& \int^{Q_2^+}_{-Q_1^+} dk^+\dfrac{\mathcal{N}\mid_{k^-=Q_2^-, k^+ = 0}}
{\left(2k^+Q_2^- -\vect{k}^2+i\epsilon\right)\left(-2k^+Q_1^- -\vect{k}^2+i\epsilon\right)}
\end{align}

Given that $|\vect{k}| \ll Q_i^+,Q_i^-$, the integrand of the $k^+$ integration in \eqref{intk-2}
is strongly peaked near the origin, and falls off rapidly before either of the two endpoints
of integration are reached. We can replace the limits of the integration by $\pm \infty$ 
without affecting the leading singularity in the integral. This allows us to perform the
$k^+$ integral using contour integration, closing in the lower half plane and picking up
the pole at $k^+ = \vect{k}^2/(2Q_2^-)$:
\begin{align} \label{intk+1}
L_{DPS} \simeq& \dfrac{(2\pi i)^2}{4(Q_1^++Q_2^+)(Q_1^-+Q_2^-)}\int_{|\vect{k}|\ll Q_i^+,Q_i^-} 
\dfrac{d^{d-2}\vect{k} \ \mathcal{N}\mid_{k^-=Q_2^-,k^+= 0}}{(\vect{k-Q_2})^2\vect{k}^2}
\end{align}

 Noticing that $4(Q_1^++Q_2^+)(Q_1^-+Q_2^-)$ is simply equal to
$2s$, we obtain a compact expression for the leading $\vect{Q}_2$ singularity in $L$:
\begin{align} \label{HsiangGeneral}
L_{DPS} \simeq& \dfrac{(2\pi i)^2}{2s}\int_{|\vect{k}|\ll Q_i^+,Q_i^-} 
\dfrac{d^{d-2}\vect{k} \; \mathcal{N}\mid_{k^-=Q_2^-,k^+= 0}}{(\vect{k-Q_2})^2\vect{k}^2}
\end{align}
The same result may be obtained by closing the $k^+$ integration in the upper half plane.

Using \eqref{HsiangGeneral}, we can reproduce the leading low $\vect{Q}_2$ behaviour of all
of the DPS boxes described in the previous section. To obtain the 4D scalar box
result \eqref{Ninh4D} , we set $\mathcal{N} = 1$ and $d = 4+2\varepsilon$ (note that, just as in
section \ref{sec:BoxSingularities}, we must perform the calculation here in slightly more than 4 dimensions
to regulate the collinear divergence in the loop integral):
\begin{align} \label{Hsiang4D}
L_{DPS,\phi,4D} =& \dfrac{(2\pi i)^2}{2s}\int_{|\vect{k}|\ll Q_i^+,Q_i^-} 
\dfrac{d^{2+2\epsilon}\vect{k}}{(\vect{k-Q_2})^2\vect{k}^2}
\\ \nonumber
\simeq& \dfrac{(2\pi i)^2}{2s}\int 
\dfrac{d^{2+2\varepsilon}\vect{k}}{(\vect{k-Q_2})^2\vect{k}^2}
\\ \nonumber
=& -\dfrac{4\pi^3}{s\varepsilon \vect{Q_2}^2}
\end{align}

We can expand the domain of integration to infinity because the integrand is strongly
peaked at $\vect{k}=\vect{0}$ when $\vect{Q_2}$ is small. The usual method of Feynman
parameters has been used to arrive at the final result.

The form of the integrand in \eqref{Hsiang4D} makes particularly clear the interplay
between the collinear and DPS divergences in the 4D scalar box integral, and their 
origins. When $\vect{Q_2} \neq 0$, there are effectively two distinct poles in the $\vect{k}$
integration, producing an overall logarithmic divergence in the integral. One of these
is associated with the loop particles on the right hand side of figure \ref{fig:genericbox} becoming 
collinear ($\vect{k} = 0$) whilst the other is associated with the particles on the
left hand side becoming collinear ($\vect{k-Q_2}=0$). As $Q_2$ is reduced to zero, the
two poles merge to form a double pole and the divergence in the integral becomes
stronger (single inverse power rather than logarithmic). The double pole is now
associated with all of the particles in the loop becoming collinear, and the
stronger divergence in the integral is precisely the DPS divergence.

Let us next consider the 6D scalar box:
\begin{align} \label{Hsiang6D}
L_{DPS,\phi,6D} =& \dfrac{(2\pi i)^2}{2s}\int_{|\vect{k}|\ll Q_i^+,Q_i^-} 
\dfrac{d^{4}\vect{k}}{(\vect{k-Q_2})^2\vect{k}^2}
\end{align}

From this expression, we can clearly see that the 6D scalar box does not possess
any collinear divergences. 

In the 6D case, we cannot straightforwardly apply the method of extending the
integration region to infinity that we used in the 4D case. The reason for this
is that the integrand no longer falls away sufficiently quickly as $|\vect{k}| \to \infty$,
and we would get infinity if we extended the integration region. 

We could evaluate the integral \eqref{Hsiang6D} by imposing a sharp cutoff $\Lambda$ on
the integration over $|\vect{k}|$, where $|\vect{Q_2}| \ll \Lambda \ll Q_i^+,Q_i^-$.
However, in practical terms it is simpler to use dimensional regularisation to extract
the leading low $\vect{Q_2}$ behaviour in $L_{DPS,\phi,6D}$. We evaluate the integral
in $6-2\varepsilon$ dimensions -- this allows us to extend the integration region to
infinity without getting an infinite result. The previously infinite contribution from
the high $\vect{k}$ end of the integral now manifests itself as a term containing a single 
pole in $1/\varepsilon$. This can simply be dropped, since we only want the contribution
from the low $\vect{k}$ end of the integral. Indeed, we discard every term except for 
the most singular term in $\vect{Q_2}$. As is typical, the dimensional
regularisation approach is conceptually more difficult to handle -- but it produces
the same result as the sharp cut-off for the leading singularity in $\vect{Q_2}$.

Applying the method, we obtain a result which agrees with \eqref{Ninh6D}:
\begin{align} \label{Hsiang6Ddimreg}
L_{DPS,\phi,6D} =& \dfrac{(2\pi i)^2}{2s}\int
\dfrac{d^{4-2\varepsilon}\vect{k}}{(\vect{k-Q_2})^2\vect{k}^2} - \text{UV pole}
\simeq  \dfrac{2\pi^4\log(\vect{Q_2}^2)}{s}
\end{align}

It turns out that we must use the dimensional regularisation method to evaluate 
the integral \eqref{HsiangGeneral} for the Yukawa $\phi\phi \to \phi^*\phi^*$, 
SM $gg \to HH$ and SM $gg \to ZZ$ cases as well. We evaluated the numerator factors for
these integrals using \vb{FORM} \cite{2000math.ph..10025V}.
The results of the calculations are listed below -- for the $gg \to HH$ and $gg \to ZZ$
cases, we only list results for the helicity amplitudes which give a nonzero result 
for $L_{DPS}$ (i.e. are divergent in the limit $\vect{Q_2} \to 0$):

\vspace{2mm}

Yukawa $g_sg_s \to g_s^*g_s^*$:
\begin{align} \label{HsiangYukawa}
L_{DPS} =& \dfrac{(2\pi i)^2}{2s}\int
\dfrac{d^{2-2\varepsilon}\vect{k}
\mathrm{Tr}[(Q_2^-\gamma^+-\slashed{\vect{k}})(Q_1^-\gamma^++\slashed{\vect{k}})(Q_1^+\gamma^--\slashed{\vect{k}}+\slashed{\vect{Q}}_2)(Q_2^+\gamma^-+\slashed{\vect{k}}-\slashed{\vect{Q}}_2)]
}{(\vect{k-Q_2})^2\vect{k}^2} 
\\ \nonumber
&- \text{UV pole}
\\ \nonumber
\simeq& 4\pi^3\log(\vect{Q_2}^2)
\end{align}

\vspace{2mm}

SM $gg \to HH$:
\begin{align} \label{HsiangHiggs}
L_{DPS}(++) = L_{DPS}(--) \simeq& -\dfrac{8 M_H^2 \pi^3 \log(\vect{Q_2}^2)}{s}
\end{align}

\vspace{2mm}

SM $gg \to ZZ$:
\begin{align} \label{HsiangZboson}
L_{DPS}(++++) &= L_{DPS} = (----) = \dfrac{4\pi^3\left[s-2M_Z^2+s\sqrt{1-4M_Z^2/s}\right]\log(\vect{Q_2}^2)}{s}
\\
L_{DPS}(++--) &= L_{DPS} = (--++) = \dfrac{4\pi^3\left[s-2M_Z^2-s\sqrt{1-4M_Z^2/s}\right]\log(\vect{Q_2}^2)}{s}
\end{align}

The results \eqref{HsiangYukawa}-\eqref{HsiangZboson} agree with those presented in section \ref{sec:BoxSingularities}, both in terms
of dependence on kinematical variables, and in terms of the numerical prefactors.

In the numerator factor of each of these box integrals, the terms with the smallest number of powers of $\vect{k}$
and/or $\vect{k-Q_2}$ are proportional to $\vect{k}\cdot(\vect{k-Q_2}$) -- the coefficients of the terms with 
lower powers of $\vect{k}$ and/or $\vect{k-Q_2}$ are all zero. A consequence of the numerators having this 
structure is that the leading $\vect{Q_2}$ singularity in each amplitude is demoted from a single inverse 
power of $\vect{Q_2}$ to a logarithm. A further consequence is that the amplitudes are free from collinear
singularities. 

Let us consider the broad features of the method that we have just introduced for isolating
the low $\vect{Q_2}$ singularity of a box. It consists of performing two sequential integrations
over the full real axis, picking up the contribution from exactly one pole each time, and then performing
the integration over $\vect{k}$. Picking up the contribution for a particular pole is equivalent to 
replacing the denominator factor corresponding to the pole by a delta function ($\times 2\pi i$). 
Essentially, our method is equivalent to replacing the $k^2$ and $(k-Q_2)^2$ factors in the denominator
by $2\pi i\delta(k^2)$ and $2\pi i\delta[(k-Q_2)^2]$ respectively, before multiplying by $-1$. We then 
neglect all the numerator terms during the $\vect{k}$ integration other than the ones with the lowest 
powers of $\vect{k}$ and/or $\vect{k-Q_2}$, such that we pick up the leading singularity in $\vect{Q_2}$. It 
is not hard to show the equivalence explicitly:
\begin{align} \label{Cutequiv}
&-(2\pi i)^2\int d^dk \dfrac{\mathcal{N}}{[(p_1+k-Q_2)^2+i\epsilon][(p_2-k)^2+i\epsilon]}\delta(k^2)\delta((k-Q_2)^2)
\\ \nonumber
=& -(2\pi i)^2\int d^{d-2}\vect{k}dk^+dk^- \dfrac{\mathcal{N}}{[(p_1+k-Q_2)^2+i\epsilon][(p_2-k)^2+i\epsilon]}
\\ \nonumber
&\times \dfrac{1}{2k^-}\delta\left(k^+ = \dfrac{\vect{k}^2}{2k^-}\right)\dfrac{1}{2[(k^+-Q_2^+)-(k^--Q_2^-)k^+/k^-]}\delta\left(k^- = Q_2^-+\dfrac{(\vect{k-Q_2})^2}{2(k^+-Q_2^+)}\right)
\\ \nonumber
=& (2\pi i)^2\int d^{d-2}\vect{k}dk^+dk^- \dfrac{\mathcal{N}}{2s\vect{k}^2(\vect{k-Q_2})^2}\delta\left(k^+ = \dfrac{\vect{k}^2}{2k^-}\right) \delta\left(k^- = Q_2^-+\dfrac{(\vect{k-Q_2})^2}{2(k^+-Q_2^+)}\right)
\\ \nonumber
&  + \text{higher order in $\vect{Q_2}$}
\\ \nonumber
=& \dfrac{(2\pi i)^2}{2s}\int d^{d-2}\vect{k} \dfrac{\mathcal{N}_{k^+=0,k^-=Q_2^-}}{\vect{k}^2(\vect{k-Q_2})^2} + \text{higher order in $\vect{Q_2}$}
\end{align}

\begin{figure}
\centering
\includegraphics[scale=0.6]{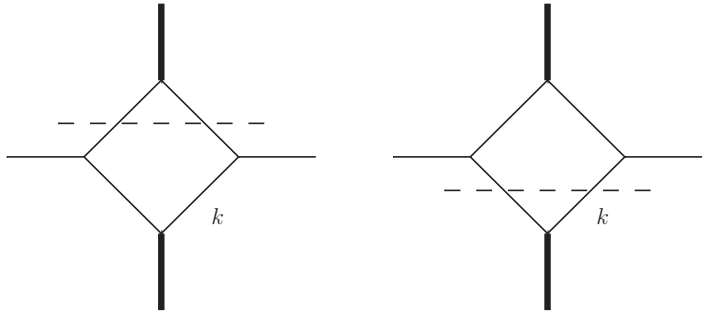}
\caption{\label{fig:cutboxes} The two cuts of the crossed box that can give rise to on-shell particles.}
\end{figure}

It should not be a surprise that the leading $\vect{Q_2}$ singularity of a box can be obtained by replacing
the $k^2$ and $(k-Q_2)^2$ denominator factors by delta functions. Notice that the leading $\vect{Q_2}$ 
singularity always appears in the real part of $L$. This corresponds to the imaginary part of a box
amplitude $\mathcal{M}$ since $L$ is always multiplied by $-i$ (along with vertex factors etc.) to make an
amplitude. But we can obtain the imaginary part of an amplitude by using the Cutkosky rules 
\cite{Cutkosky:1960sp, Veltman:1994Ve}. Thus, twice the real part of $L$ is given by minus the sum over all cuts 
for which the cut propagators may be put on shell (the minus comes from the fact that $\mathcal{M}
 \propto -iL$). There are two such cuts for the box diagram, which we have drawn in figure \ref{fig:cutboxes}. They 
give equivalent contributions in the small $\vect{k}$ and $\vect{k-Q_2}$ limit, with both contributions
being equal to minus \eqref{Cutequiv}. Putting everything together, we see that the Cutkosky rules
predict that the leading $\vect{Q_2}$ singularity in the real part of $L$ (= the leading singularity 
in $L$) is given by \eqref{Cutequiv}. 

Inserting the values for $d$ and $\mathcal{N}$ for the 6D scalar, 4D Yukawa, and 4D Standard Model crossed 
boxes into \eqref{Cutequiv}, one might get the impression that the real parts of the loop integrals $L$ for 
these boxes are all ultraviolet divergent. It is well-known that an ultraviolet divergence exists in the 
4D Yukawa and Standard Model crossed boxes -- however, this occurs in the imaginary part, as can be verified
by examining the loop integral expressions in the large $k$ limit, and remembering that that a factor of $i$
appears during Wick rotation. What we have not written down explicitly in \eqref{Cutequiv}, but is easy to
show, is that for large $\vect{k}$ both delta functions cannot be satisfied simultaneously. Therefore the 
integral is effectively cut off at large $\vect{k}$ and the real part of $L$ for any crossed box is UV
finite. The appropriate integration region in $\vect{k}$ space for the real part of $L$ is an ellipse with 
the foci at $\vect{0}$ and $\vect{Q_2}$ and semi-minor axis length $M/2$. This approximates to a circle of
radius $M/2$ centred at the origin when $\vect{Q_2}$ is small.

\begin{figure}
\centering
\includegraphics[scale=0.59]{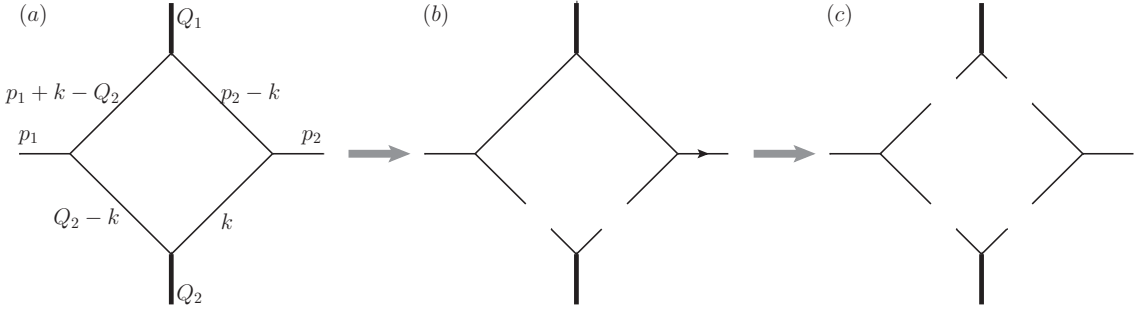}
\caption{\label{fig:boxdecomp} Decomposition of the box integrand.}
\end{figure}

In the presence of the two delta functions, the remainder of the integrand in \eqref{Cutequiv} can be
decomposed into two factors, corresponding to the two Feynman diagrams of figure \ref{fig:boxdecomp}(b). 
Given that the lines with momentum $p_1+k-Q_2$ and $p_2-k$ are almost on shell when $\vect{k}$ and $\vect{k-Q_2}$
are small, we can use completeness relations to further decompose the upper diagram of figure 
\ref{fig:boxdecomp}(b) into three smaller diagrams (divided by two propagator factors) -- see figure 
\ref{fig:boxdecomp}(c). This procedure is very similar to, say, the textbook decomposition of the matrix 
element for $e^-X \to \gamma Y$ into $e^- \to \gamma e^-$, $e^-X \to Y$ (divided by a propagator factor) in the collinear limit
(see Chapter 17 of \cite{Peskin:1995ev}).

Applying the decomposition of figure \ref{fig:boxdecomp}, the leading low $\vect{Q_2}$ divergence of
a general crossed box may be written as follows (recall that our labelling conventions are given in figure 
\ref{fig:genericbox}):
\begin{align} \label{boxdecomp}
L_{DPS}(\lambda_1\lambda_2\mu_1\mu_2) =& \sum_{s_i, L_i}\int d^d k \delta(k^2)\delta((k-Q_2)^2) 
\Phi_{b \to L_2L_3}^{\lambda_2 \to s_2s_3}(p_2;p_2-k,k) 
\\ \nonumber
\times& \Phi_{a \to L_1L_4}^{\lambda_1 \to s_1s_4}(p_1;p_1+k-Q_2,Q_2-k)
\mathcal{M}^{s_3s_4 \to \mu_2}_{L_3L_4 \to B} (k,Q_2-k;Q_2)
\\ \nonumber
\times& \mathcal{M}^{s_1s_2 \to \mu_1}_{L_1L_2 \to A} (p_1+k-Q_2,p_2-k;Q_1)
\end{align}

$\Phi_{a \to bc}^{\lambda \to s_1s_2}$ is essentially the light-cone wavefunction to find the pair $bc$
with helicities $s_1s_2$ inside the particle $a$ with helicity $\lambda$ \cite{Harindranath:1998pd}. Each
of these functions in \eqref{boxdecomp} is composed from three ingredients -- the matrix element from the
relevant Feynman diagram in figure \ref{fig:boxdecomp}(c), the propagator factor nearest to this diagram, and
one further factor $R$. The last factor is equal to the square rooted ratio of the collinear momentum
fractions of the upper and lower outgoing particles in the relevant Feynman diagram. In the spirit of 
\cite{Peskin:1995ev}, the matrix elements in the $\Phi$ factors of \eqref{boxdecomp} should be evaluated 
using the following approximate expressions for the loop vectors:
\begin{align}
k = Q_2^-n + \vect{k};& \qquad Q_2-k = Q_2^+p - (\vect{k-Q_2})
\\ \nonumber
p_1+k-Q_2 = Q_1^+p + (\vect{k-Q_2});& \qquad p_2-k = Q_1^-n -\vect{k}
\end{align}

$\mathcal{M}^{s_3s_4 \to \mu_2}_{L_3L_4 \to B}$ is the matrix element for the `hard process' in which the pair $L_3L_4$
with helicities $s_3s_4$ interact to make particle $B$ with helicity $\mu_2$. Given that we are only interested in
extracting the leading $\vect{Q_2}$ singularity of $L_{DPS}$, it is actually acceptable to evaluate this ingredient of
\eqref{boxdecomp} with all transverse momenta set to zero.

It should be pointed out that the formula \eqref{boxdecomp} only strictly applies when the masses of $A$ and $B$ 
are equal. The reason for this is that to introduce the $R$ factors which are a part of the $\Phi$ functions into
the box integrand, we have used the fact that $R$ for the left hand $\Phi$ is the reciprocal of the $R$ for the
right hand $\Phi$. Then we can introduce the $R$ functions via $1 = R_{left}R_{right}$. This relation only
actually holds when $M_A = M_B$. In the more general case in which $M_A$ is not necessarily equal to $M_B$, there
will be a prefactor equal to  $M_B/M_A$ in front of \eqref{boxdecomp}.

From our experience of the QCD light cone wavefunction, we can say that $\Phi_{a \to bc}^{\lambda \to s_1s_2}$
will in general factorise into two parts, one of which is only dependent on the transverse momentum of $b$ relative
to $a$, and the other of which is only dependent on the collinear fraction of the momentum of $a$ that is carried by $b$.
So, for $\Phi_{a \to L_1L_4}$ for example:
\begin{equation} \label{LCdecomp}
\Phi_{a \to L_1L_4}^{\lambda_1 \to s_1s_4}(p_1;p_1+k-Q_2,Q_2-k) = X_{a \to L_1L_4}^{\lambda_1 \to s_1s_4}\left(\dfrac{Q_1^+}{Q_1^++Q_2^+}\right)
K_{a \to L_1L_4}^{\lambda_1 \to s_1s_4}(\vect{k-Q_2})
\end{equation}

The factor $X$ in \eqref{LCdecomp} can be interpreted as the square root of the real splitting part of a helicity dependent splitting function.
In scalar field theory, the function $K_{\phi \to \phi\phi}(\vect{k})$ is simply the $1/\vect{k}^2$ coming from the 
propagator factor since the splitting matrix element is proportional to $1$ in this theory. On the other hand, all of
the $K$ functions for QCD, QED and scalar gluon theory only diverge like $1/\vect{k}$ for small $\vect{k}$. The reason
for this is that all of the $1 \to 2$ splittings in these theories are forbidden in the absolute collinear limit due
to nonconservation of $J_z$. This means that the splitting matrix elements must all be proportional to $\vect{k}$,
which goes together with the $1/\vect{k}^2$ from the propagator factor to produce a $1/\vect{k}$ dependence for $K$.
In QED/QCD, $J_z$ is not conserved for the $g/\gamma \to q\bar{q}$ collinear splitting because the initial state must 
have helicity $\pm 1$, and the structure of the theory forces the quark and antiquark in the final state to have opposite 
helicities (i.e. total $J_z = 0$). In scalar gluon theory, $J_z$ is not conserved for the $g_s \to q\bar{q}$ collinear splitting
because the initial state has helicity 0, and the structure of the theory in this case forces the outgoing fermions to both have the
same helicity (i.e. total $J_z = \pm 1$). 

Recall that we can consider $\mathcal{M}$ as being independent of $\vect{k}$, since we can calculate it in the limit
in which $\vect{k}$ is zero. Thus, ignoring $\mathcal{M}$ and the $X$ parts of $\Phi$ in \eqref{boxdecomp}, which will
only contribute to the prefactor of the leading $\vect{Q_2}$ divergence in $L_{DPS}$, we can write $L_{DPS}$ schematically
as:
\begin{align} \label{bxdcmpsch}
L_{DPS}(\lambda_1\lambda_2\mu_1\mu_2) \sim& \sum_{s_i}\int d^{d-2}\vect{k} K_{a \to L_1L_4}^{\lambda_1 \to s_1s_4}(\vect{k-Q_2})
K_{b \to L_2L_3}^{\lambda_2 \to s_2s_3}(-\vect{k}) 
\end{align}

This integral bears a strong resemblance to the integral that one would use to investigate what kind of scaling violations
a theory has in its parton distributions, which schematically is $\int d^{d-2}\vect{k} |K(\vect{k})|^2$ (if this
integral diverges logarithmically, then the theory has logarithmic scaling violations, etc.). It is reasonably clear 
from this connection that if a theory has logarithmic scaling in its parton distributions, then it will also have 
logarithmic DPS singularities in its crossed boxes. The disappearance of the $1/\vect{Q_2}^2$ singularity in the 
$gg \to ZZ$ and $gg \to HH$ boxes may be traced to the fact that the $K$ functions in QCD have numerator factors
proportional to $\vect{k}$. This in turn is caused by the vector nature of the QCD theory resulting in a vanishing of the 
matrix elements for collinear splittings, and is not directly related to the gauge nature of the theory as was suggested 
in \cite{Ninh:2008cz, Bern:2008ef}.

For the process $g \to q\bar{q}$, we present below explicit expressions for the functions $\Phi$ for all possible helicity 
configurations. Overall numerical prefactors are emitted in these expressions -- we give only the dependence on the 
transverse momentum of the quark $\vect{k}$ and the collinear fraction of the gluon's momentum that goes to the quark $x$:
\begin{align} \label{LCgqq}
 \Phi_{g\to q\bar{q}}^{+ \to +-}(x,\vect{k}) \propto x (\vect{\epsilon^+}\cdot\vect{k})/\vect{k^2}  \qquad
 \Phi_{g\to q\bar{q}}^{+ \to -+}(x,\vect{k}) \propto (1-x) (\vect{\epsilon^+}\cdot\vect{k})/\vect{k^2} 
\\\nonumber
 \Phi_{g\to q\bar{q}}^{- \to -+}(x,\vect{k}) \propto x (\vect{\epsilon^-}\cdot\vect{k})/\vect{k^2}  \qquad
 \Phi_{g\to q\bar{q}}^{- \to +-}(x,\vect{k}) \propto (1-x) (\vect{\epsilon^-}\cdot\vect{k})/\vect{k^2} 
\end{align}

$\vect{\epsilon^+}$ ($\vect{\epsilon^-}$) is the transverse part of the polarisation vector with 
positive (negative) helicity along the gluon direction.

The unpolarised and polarised $g \to q$ splitting functions divided by $|\vect{k}|$ are formed from appropriate
linear combinations of the mod squares of these $\Phi$ functions:

\begin{align}
\sum_{\lambda,s_1,s_2}|\Phi^{\lambda \to s_1s_2}_{g \to q\bar{q}}|^2 \propto& \dfrac{x^2 + (1-x)^2}{\vect{k}^2} \propto \dfrac{P_{qg}(x)}{\vect{k}^2}
\\ 
\sum_{\lambda,s_1,s_2}\lambda\dfrac{s_1}{|s_1|} |\Phi^{\lambda \to s_1s_2}_{g \to q\bar{q}}|^2 \propto& \dfrac{x^2 - (1-x)^2}{\vect{k}^2}
\propto \dfrac{\Delta P_{qg}(x)}{\vect{k}^2}
\end{align}

Let us consider the box integral \eqref{boxdecomp} for the process $gg \to AB$ with arbitrary final states, and ignore
the $\mathcal{M}$ and $X$ functions in \eqref{boxdecomp} which do not depend on $\vect{k}$. If the two initial state
gluons have the same helicity, then in the limit $\vect{Q_2} = 0$ this integral looks like $\int d^2 \vect{k} 
(\vect{\epsilon^{+z}} \cdot \vect{k}) (\vect{\epsilon^{-z}} \cdot \vect{k}) / \vect{k}^4$ [$\vect{\epsilon^{+z}}
= (1,i)$ and $\vect{\epsilon^{-z}} = (1,-i)$]. This is logarithmically divergent. On the other hand, when the 
gluon helicities are opposite, we get $\int d^2 \vect{k} (\vect{\epsilon^{\pm z}} \cdot \vect{k}) 
(\vect{\epsilon^{\pm z}} \cdot \vect{k}) / \vect{k}^4$ which evaluates to zero. Thus, the $gg \to AB$ crossed
box will not contain a logarithmic DPS singularity if the initial state gluons have opposite helicities. It is
important to emphasise that this statement is totally independent of the final states $AB$. Note this general
rule is obeyed for the case of the $gg \to ZZ$ and $gg \to HH$ crossed boxes -- see \eqref{HsiangHiggs} and
\eqref{HsiangZboson}.

The physical explanation for this phenomenon is as follows. The structure of the QCD theory forces the 
$q\bar{q}q\bar{q}$ intermediate state in the crossed box process $ gg \to q\bar{q}q\bar{q} \to AB$ (which
is essentially real in the collinear limit) to have total $J_z = 0$ in the collinear limit. Then, if the
initial state gluons have opposing helicities $J_z = \pm 2$, there is an issue with total $J_z$
nonconservation aside from local $J_z$ nonconservation at each $g \to q\bar{q}$ vertex. This manifests
itself as a further suppression of the $ gg \to AB$ box integral numerator in the limit 
$\vect{k}, \vect{Q_2} \to 0$, which makes the integral convergent.

If the final state particles $AB$ have spin, then there is one further way in which a $gg \to AB$ 
crossed box can become convergent in the limit $\vect{Q_2} = 0$ contrary to naive expectations. If the
helicities of $A$ and $B$ are such that there is no assignment of helicities to the internal lines which
simultaneously conserves helicity at the $g \to q\bar{q}$ vertices, and conserves $J_z$ at the $q\bar{q} \to A$
and $q\bar{q} \to B$ vertices in the collinear limit, then the crossed box integral will not contain
a DPS singularity. The extra numerator suppression in the limit $\vect{k}, \vect{Q_2} \to 0$ comes from
one or both of the factors $\mathcal{M}$ in this case. This rule can be seen to hold in the case
$gg \to ZZ$.

We can make some sense of the prefactors in \eqref{HsiangYukawa} - \eqref{HsiangZboson} in terms of
products of square roots of helicity dependent splitting functions using our decomposition \eqref{boxdecomp}.
Where the factors $\mathcal{M}$ are nonzero, they can only be proportional to $M$ regardless of the
final state. We also find $\int d^4 k \delta(k^2)\delta((k-Q_2)^2) \propto \int d^2 \vect{k}/M^2$ for small $\vect{k},
\vect{Q_2}$. Taking $d=4$ in $L_{DPS}$:
\begin{align} \label{boxdecomp2}
L_{DPS}(\lambda_1\lambda_2\mu_1\mu_2) \propto& \sum_{s_i, L_i} X_{g \to L_1L_4}^{\lambda_1 \to s_1s_4}\left(x\right)
X_{g \to L_2L_3}^{\lambda_2 \to s_2s_3}\left(1-y\right) \mathcal{\delta}^{s_1s_2 \to \mu_1}_{L_1L_2 \to A}
\mathcal{\delta}^{s_3s_4 \to \mu_2}_{L_3L_4 \to B} 
\\ \nonumber
\times& \int d^2 \vect{k} K_{g \to L_1L_4}^{\lambda_1 \to s_1s_4}(\vect{k} - \vect{Q_2})
K_{g \to L_2L_3}^{\lambda_2 \to s_2s_3}(-\vect{k}) 
\end{align}
where $\mathcal{\delta}^{s_1s_2 \to \mu_1}_{L_1L_2 \to A}$ is simply a function that is equal to $1$ if $J_z$ is conserved
in the `hard process' producing final state $A$, and zero otherwise. Here $x$ is defined to be equal to $Q_1^+/(Q_1^++Q_2^+)$
and $y = Q_2^-/(Q_1^-+Q_2^-)$. Since we have taken the masses of $A$ and $B$ equal and work in the centre of mass frame,
$Q_1^+ = Q_2^-, Q_1^- = Q_2^+$, and $y=x$.

It is clear that the second line of \eqref{boxdecomp2} provides the factor of $\log(\vect{Q_2}^2)$, whilst the first line 
provides the prefactor that depends on $M$ and $s$. Without loss of generality, let us take $Q_1^+ > Q_1^-, Q_2^- > Q_2^+$
-- i.e. we take $A$ to be the final state particle that travels along the $+z$ axis in the collinear limit.

\begin{figure}
\centering
\includegraphics[scale=0.6]{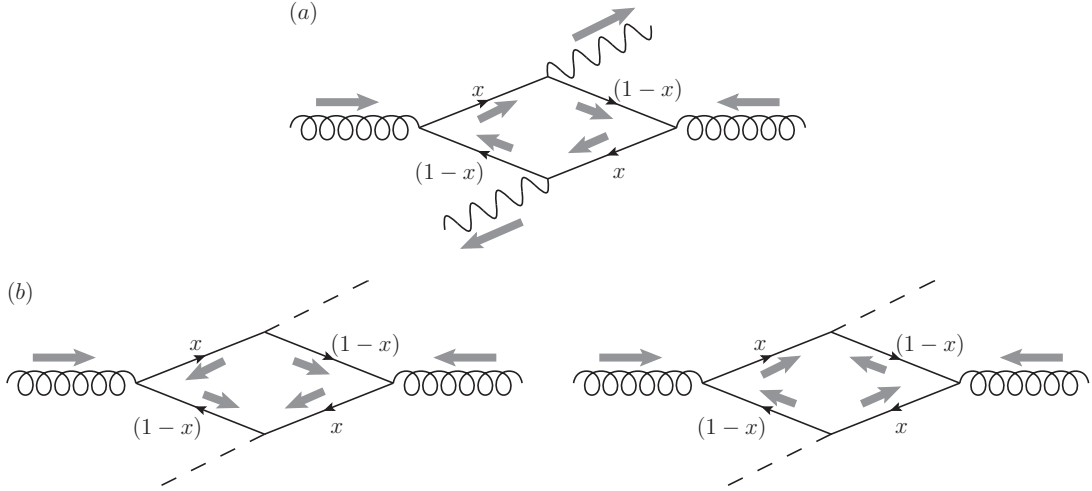}
\caption{\label{fig:helicitybox} Possible configurations of internal helicity for the $gg \to AB$ crossed box in the collinear
limit, where we have taken $Q_1^+ > Q_1^-, Q_2^- > Q_2^+$, both initial state gluons have positive helicity, and the final 
state is $(a)$ a pair of Z bosons with positive helicity $(b)$ a pair of Higgs bosons. Note that the permitted internal helicity
configurations would be the same for both $(a)$ and $(b)$ if the helicities of the gluons were both negative instead.}
\end{figure}

Consider the $++++$ helicity configuration for the $gg \to ZZ$ box. In the collinear limit there is only one possible
assignation of helicities to the internal lines which is permitted by the structure of the theory and conserves $J_z$ at the 
hard processes. This is presented in figure \ref{fig:helicitybox}(a). In this diagram, the internal lines with positive helicity both have 
collinear momentum fraction of parent equal to $x$. Bearing this in mind, and using the formulae \eqref{LCgqq}, we see that
the prefactor of the $++++$ crossed box must be proportional to $x^2$, which in turn is proportional to 
$(s-2M_Z^2+s\sqrt{1-4M_Z^2/s})/s$.

This process can be repeated for all other processes and helicity configurations. The internal helicity configuration is the
same for the $gg \to ZZ$ $--++$ process -- looking at \eqref{LCgqq} we can then clearly see that the prefactor must be
proportional to $(1-x)^2 \propto (s-2M_Z^2-s\sqrt{1-4M_Z^2/s})/s$. For the $gg \to HH$ diagram, the two possible arrangements
of internal helicities are always the same regardless of the gluon helicities (figure \ref{fig:helicitybox}(b)). For both of
these arrangements, the internal lines with the same helicity always have complementary momentum fractions (this is different
from the $gg \to ZZ$ case, in which internal lines with the same helicity always have the same momentum fraction). 
As a result of this, the prefactor for the $gg \to HH$ process is 
proportional to $x(1-x) \propto M_H^2/s$. Finally, the $g_s \to q\bar{q}$ light cone wavefunctions do not contain any 
dependence on $x$ (a consequence of this being that that $P_{qg_s}$ does not depend on $x$ \cite{Tosa:1979id}), so the 
prefactor of the $g_sg_s \to g_s^*g_s^*$ crossed box does not contain any dependence on $s$ or $M$.

Actually, we can also justify the prefactor for the 6D scalar box \eqref{Hsiang6Ddimreg} using the framework of
\eqref{boxdecomp}. For the 6D scalar case $\mathcal{M}$ is now independent of $M$, so the equivalent expression to
\eqref{boxdecomp2} in this case has a prefactor of $1/M^2$, besides having $d^2\vect{k}$ replaced by $d^4\vect{k}$
and all helicity labels removed. As there are no complications involving spin for 6D scalar theory, we can 
straightforwardly associate the function $X_{\phi \to \phi\phi}(x)$ with the square root of the real splitting part
of the $\phi \to \phi$ splitting function in 6D $\phi^3$ theory, which is given by $P_{\phi\phi}(x) \propto x(1-x)$ \cite{Konishi:1979cb}.
Putting everything together, we find that the prefactor for the crossed box in 6D $\phi^3$ theory is proportional to
$x(1-x)/M^2 = 1/s$.

It is not hard to show that equation \eqref{boxdecomp2} continues to hold even when the masses of $A$ and $B$ are not
equal, although one has to bear in mind that $x$ is not necessarily equal to $y$ in general. It is easy to use this
expression (or the scalar 4D/6D equivalent) to generalise the results \eqref{Hsiang4D} - \eqref{HsiangZboson} to 
arbitrary masses for $A$ and $B$. We only write down one of these generalisations here -- $gg \to AB$, where $A$
and $B$ are scalars -- and leave the others as exercises for the reader. The $\log{\vect{Q_2}^2}$ prefactors of the
DPS divergent graphs in this case ($++$ and $--$) are identical and proportional to $x(1-y) + y(1-x) \propto (M_A^2+M_B^2)/s$. The two terms
in this result are associated with the two diagrams of figure \ref{fig:helicitybox}(b).

Let us consider the part of the $pp \to AB + X$ cross section associated with two gluons splitting almost collinearly
into quark and antiquark pairs, and then these pairs interacting to form $A$ and $B$. Suppressing helicity and colour
indices:
\begin{align} \label{ggDPSXsec}
\sigma_{pp \to gg \to AB + X,DPS}(s) =& \int dX d\bar{X} f_g(X) f_g(\bar{X}) \hat{\sigma}_{gg \to AB, DPS}(\hat{s} = sX\bar{X})
\\ \label{ggDPSXsecpartonlevel}
\hat{\sigma}_{gg \to AB, DPS}(s) \propto& \dfrac{1}{s} \int d^4q_1 d^4q_2 \delta(q_1^2-M^2) \delta(q_2^2 - M^2) 
\delta^{(4)} (q_1+q_2-p_1-p_2) 
\\ \nonumber
&\times |L_{DPS, gg \to AB}|^2
\end{align}

By decomposing $L_{DPS}$ according to \eqref{boxdecomp} and \eqref{LCdecomp}, and then making a few substitutions for the
integration variables in \eqref{ggDPSXsec},\eqref{ggDPSXsecpartonlevel}, one finds that one can bring \eqref{ggDPSXsec}
into the form of a double parton scattering cross section expressed in terms of the two-parton GPDs $\Gamma$ of
\cite{Diehl:2011tt}:
\begin{align} \label{ggDPSXsec2}
\sigma_{pp \to gg \to AB + X,DPS} \propto & \int \prod_{i=1}^{2}dx_id\bar{x}_i \hat{\sigma}_{q\bar{q} \to A}(\hat{s} = x_1\bar{x}_1s) 
\hat{\sigma}_{q\bar{q} \to B} (\hat{s} = x_2\bar{x}_2s)
\\ \nonumber
\times & \int \dfrac{d^2\vect{r}}{(2\pi)^2} 
\Gamma_{q\bar{q}}|_{g \to q\bar{q}}(x_1,x_2, \vect{r}) \Gamma_{q\bar{q}}|_{g \to q\bar{q}}(\bar{x}_1,\bar{x}_2, -\vect{r}) 
\end{align}

The $\Gamma$ factors in \eqref{ggDPSXsec2} are not the full two-parton GPDs -- rather they are the contributions to these
objects coming from a $g \to q\bar{q}$ splitting, as defined in section 12 of \cite{Diehl:2011tt}. Equation \eqref{ggDPSXsec2}
is somewhat schematic, in that our full result (and the full expression for the DPS cross section in \cite{Diehl:2011tt}) is actually 
a sum over terms containing helicity-dependent cross sections and helicity-dependent two-parton GPDs that are either both diagonal
or both off-diagonal in helicity space. The same formula \eqref{ggDPSXsec2} is obtained for the close-to-collinear part of the 
$pp \to gg \to AB+X$ cross section if the masses of $A$ and $B$ are not equal.

Thus, it appears that the close-to-collinear part of the $pp \to gg \to AB +X$ cross section can be described using double 
parton scattering formulae involving two-parton GPDs. This is an interesting result that is not obvious from looking at the 
basic expression for the box integral \eqref{genericbox}.

The transverse momentum integrals in \eqref{ggDPSXsec2} are not divergent at the infra-red end (note that there is actually
one transverse momentum integration `hidden' in each 2pGPD factor of \eqref{ggDPSXsec2}). This is because the DPS singularity 
in SM crossed boxes is integrable. If one extends all of the transverse momentum integrations to infinity, the result is 
ultraviolet divergent -- but one might argue that it is not reasonable to do such a thing. We have seen that \eqref{ggDPSXsec2}
corresponds to the contribution to the cross section from the imaginary part of the $gg \to AB$ crossed box matrix element, which 
is UV finite. The $\vect{r}$ integral and two transverse momentum integrals inside the $\Gamma$ factors of \eqref{ggDPSXsec2} 
can be mapped via a linear change of variables to the $\vect{Q_2}$ integral and two transverse momentum integrals inside the 
$L_{DPS}$ factors of \eqref{ggDPSXsecpartonlevel}. All of the transverse momentum integrations \eqref{ggDPSXsecpartonlevel} 
are naturally cut off at finite values, implying that the ones in \eqref{ggDPSXsec2} should be also.

We would suggest that the key property of SM crossed boxes, which is correctly anticipated by the two-parton GPD framework,
is that they give rise to a cross section which is infrared finite. Thus there are no infra-red singularities in these objects
which need to be absorbed into double parton distributions -- they should be regarded entirely as a contribution to the 
single parton scattering process.

\section{DPS singularity in Loops with More Than Four Legs} \label{sec:MoreLegs}

\begin{figure}
\centering
\includegraphics[scale=0.6]{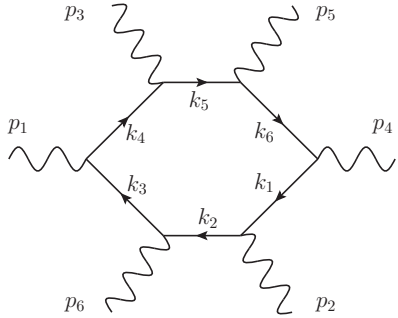}
\caption{\label{fig:6photon} The six-photon loop diagram.}
\end{figure}

An example of a loop diagram with more than four legs which contains a DPS singularity is
the six-photon amplitude displayed in figure \ref{fig:6photon}. For the diagram to contain a DPS singularity, 
we must take the initial state particles to be the photons with momenta $p_1$ and $p_4$, whilst
the remaining particles are in the final state. All of the external particles are taken to be 
on-shell (i.e. $p_i^2 = 0$). The DPS singularity occurs when the total transverse momentum $\vect{P_\Sigma}$ of
photons $3$ and $5$ (or equivalently $2$ and $6$) becomes zero. It is associated with the point
in the loop integration at which $k_1$ and $k_6$ become collinear with $p_4$, whilst $k_3$ and $k_4$
become collinear with $p_1$.

The first result for a six-photon helicity amplitude, summed over loop topologies, was obtained by Mahlon
\cite{Mahlon:1994dc} for the MHV helicity configuration. Since then, numerical techniques have been 
developed for performing the loop integration for arbitrary values of the external momentum and helicity
\cite{Nagy:2006xy, Ossola:2007bb, Gong:2008ww} and analytical expressions for all of the six photon
helicity amplitudes have been obtained in \cite{Binoth:2007ca, Bernicot:2007hs}. In \cite{Bern:2008ef,
Bernicot:2008nd}, the behaviour of an MHV and an NMHV helicity amplitude (in particular the latter) 
close to a DPS singular point is investigated. The helicity configuration in the MHV amplitude is 
$-++-++$ , whilst that for the NMHV amplitude is $---+++$ (the ordering of the helicities here corresponds
to the numbering of the external momenta, and all helicities are defined relative to incoming external momenta). 
Detailed plots are presented in \cite{Bern:2008ef,Bernicot:2008nd} illustrating the approach of the 
NMHV amplitude to the following phase space point satisfying $\vect{P_\Sigma} = 0$:
\begin{align} \label{BernicotConfig}
&\overrightarrow{p_2} = (-33.5,-15.9,-25.0) && \overrightarrow{p_3} = (-12.5,15.3,22.0) \\ \nonumber
&\overrightarrow{p_5} = (12.5, -15.3,-0.3)  && \overrightarrow{p_6} = (33.5,15.9,3.3)
\end{align}
The values given above for each photon are the $(x,y,z)$ components of the four momentum -- the remaining $t$
component is fixed by the on shellness condition. The momenta $\overrightarrow{p_1}$ and  $\overrightarrow{p_4}$
are taken to be along the positive and negative $z$ axis respectively.

The conclusion drawn from the plots is that the NMHV amplitude is finite at the DPS singular point, at least
for the configuration of external momenta \eqref{BernicotConfig}. It is also inferred that the MHV amplitude 
is finite at the singular point, from the fact that the amplitude does not contain any sharp structure when
the Nagy-Soper final state momentum configuration \cite{Nagy:2006xy} is rotated around the $y$ axis (with some
rotation angles corresponding to quite a close approach to the DPS singular point). It is implied in 
\cite{Bern:2008ef} that this behaviour is somewhat surprising, given that simple power counting arguments 
indicate that the amplitude should diverge at the DPS singular point as $1/\vect{P_\Sigma}^2$ (similar to
\eqref{Ninh4D}).

The loop decomposition technique developed in the last section can be very straightforwardly applied to the present
situation, to check the results of \cite{Bern:2008ef,Bernicot:2008nd} and investigate in more generality the
low $\vect{P_\Sigma}$ behaviour of the six-photon loop integral. One thing we can say straight
away, bearing in mind our experience with the box integrals and recalling that QED is a vector theory
like QCD, is that the DPS divergence in the loop integral can be no worse that the logarithm of 
$\vect{P_\Sigma}$. Thus there is certainly no danger of the $2\gamma \to 4\gamma$ cross section 
being infinite.

We can actually reproduce the results of \cite{Bern:2008ef,Bernicot:2008nd} without doing any further
calculations. The DPS singularity in a particular $-++-++$ MHV diagram will cancel when we add on all other loop
topologies which have their DPS singularity in the same place. The reason for this is that when we 
decompose all of these topologies according to \eqref{boxdecomp}, and then add all of the decomposed
integrals together, then one finds that one can extract two factors from the result which are equal
to the full tree-level matrix element for $q\bar{q} \to \gamma\gamma$ (i.e. the sum of both possible
Feynman diagrams). This is, of course, not unexpected. For the MHV helicity configuration considered
the helicity of both final state photons in these matrix elements will be the same. But it is well
known that the amplitude for a quark and an antiquark to produce two photons with the same helicity is
zero (see for example \cite{Ozeren:2005mp}). So the leading DPS singularity in the $-++-++$ MHV 
diagram goes to zero -- i.e. the amplitude is convergent.

The $---+++$ NMHV amplitude cannot contain a DPS singularity simply because the $J_z$ of the initial 
state is not equal to zero. We showed in the last section that a crossed box loop integral with two
gluons in the initial state does not contain a DPS singularity unless the total $J_z$ of the gluons
is equal to zero. This result obviously generalises to any one-loop integral that can potentially
contain a DPS singularity, and still applies when the initial state gluons are swapped for photons.
Thus, the DPS singularity in the NMHV amplitude vanishes on a diagram by diagram basis.

Note that these results serve as a generalisation of the results of \cite{Bern:2008ef,Bernicot:2008nd}
to the case of arbitrary initial and final state momenta. Aside from using our loop decomposition 
framework to do this, we can also use it to make some interesting statements about the singular
behaviour of the other NMHV and MHV amplitudes. First, we can say that no NMHV six-photon amplitude
can ever contain a logarithmic DPS singularity. The reason for this is that, however one distributes
the helicities, one always ends up either with the initial state photons having opposite helicities,
or with one of the pairs of the final state photons having the same helicity. On the other hand, there
are MHV amplitudes that do have logarithmic DPS singularities -- for example, the $+--+++$ configuration.

In section \ref{sec:intro}, we saw that according to the `dPDF framework' for calculating DPS cross sections,
we should be able to associate a $(\int d|\vect{k}|/|\vect{k}|)^2$ factor at the cross section level with
the DPS singularity of figure \ref{fig:dpsloops}. However, using the loop decomposition technique developed 
in section \ref{sec:PhysicalInvestigation}, we can readily see that no such factor may be associated with the 
DPS singularity of the loop. At most, we will be able to associate an integration of the form 
$\int d^2\vect{k} \log^2(\vect{k}^2)$ with the DPS singularity at the cross section level, which is not
infra-red divergent.

We may deduce that diagrams which have the structure of figure \ref{fig:dpsloops} should not be included in the leading logarithmic
DPS cross section, and thus that the dPDF framework of section \ref{sec:intro} is not sound theoretically. Given our discussion
of the crossed box in the last section, this result is hardly a surprise. The dPDF framework incorrectly predicts the singular
behaviour of even the simple $gg \to AB$ crossed box, and we have to use a framework in terms of two-parton GPDs (which have
an extra transverse momentum or impact parameter argument) to get the behaviour right.

At the heart of the problem with the dPDF framework is the assumption that the two parton GPD may be factorised into a 
longitudinal and a transverse piece, with the longitudinal piece being given by the dPDF object of \cite{Zinovev:1982be}.
In \cite{Diehl:2011tt}, the evolution of two-parton GPDs is carefully studied using perturbative QCD,
and it is established that the `single PDF feed' does not make any contribution to the two-parton GPD for parton pair
separations $\vect{b} \neq 0$. This means that the two-parton GPD evolves very differently for $\vect{b} = 0$ and 
$\vect{b} \neq 0$, which is inconsistent with the factorisation hypothesis. 

Finally, we must point out that even following the breakdown of the dPDF framework for describing proton-proton DPS, the
dPDF object of \cite{Zinovev:1982be} (which we also studied and provided explicit leading order forms for in \cite{Gaunt:2009re})
appears to retain validity as the integral of the (colour singlet) two-parton GPD over $\vect{b}$ \cite{Diehl:2011tt}. 
This is not probed directly in proton-proton collisions, but is probed in proton--heavy nucleus collisions (see 
\cite{Strikman:2001gz, Cattaruzza:2004qb}).

\section{Conclusions} \label{sec:conc}

In this paper, we have demonstrated that the DPS singular part of any one-loop diagram of the appropriate structure may be
simply expressed in terms of the transverse momentum integral of two light cone wavefunctions and two hard matrix elements.
An explicit derivation of this expression was given for the four-point case, but it is clear that such an expression will
continue to be applicable for larger numbers of external particles.

A naive treatment of Standard Model one-loop diagrams initiated by QED/QCD vertices connected to 
massless particles indicates that the DPS singularities in these diagrams should be of the same strength
as those in the corresponding diagrams with scalars -- i.e. $1/\vect{p_T}^2$, where $\vect{p_T}$ 
is the transverse momentum sum of all of the final state particles on one of the loop lines extending between the initial
state particles. Using our expression for the DPS singularity, we have shown that if the theory that determines the
vertices at which the initial particles attach has logarithmic scalings in its parton distributions (e.g. QED, QCD, scalar
gluon theory, 6D $\phi^3$), then the DPS singularity in the one-loop diagram cannot be stronger than a logarithm of 
$\vect{p_T}^2$. There is clearly a suppression of the numerator in the Standard Model loops at the DPS singular point 
which causes their DPS singularity to go from $1/\vect{p_T}^2$ to $\log(\vect{p_T}^2)$. This can be traced to the vector 
nature of the QED and QCD theories forcing $J_z$ nonconservation in, and therefore suppression of, the decay of a massless 
particle to a collinear pair of massless particles.

We exploited our framework to show that an arbitrary one-loop diagram initiated by gluons/photons with fermions running 
around the loop does not contain a DPS singularity if the total $J_z$ of the initial state is not zero. The physical
reason for this is that the total $J_z$ of the $f\bar{f}f\bar{f}$ intermediate state in the loop, which becomes real 
at the DPS singular point, is constrained to have $J_z = 0$ at the DPS singular point by the structure of QED/QCD.
If initial $J_z \neq 0$ there is then an issue of total $J_z$ nonconservation (aside from local $J_z$ nonconservation
at each vertex), which suppresses the loop numerator further at the DPS singular point and completely removes the
DPS singularity. The DPS singularity in a given diagram will also disappear if one or both of the hard matrix elements 
happen to vanish in the limit of collinear, on-shell initial state particles.

These general principles were applied to explain why the $gg \to ZZ$ and $gg \to HH$ box integrals only contain 
logarithmic DPS divergences for certain configurations of the external helicity. In both cases, a necessary 
condition for the box to have a DPS divergence is that the gluons should have the same helicity (ensuring total
$J_z=0$). In the $gg \to ZZ$ case, the $Z$ bosons must have the same helicity otherwise there is no configuration
of internal helicity which ensures $J_z$ conservation at both $q\bar{q} \to Z$ vertices in the collinear limit,
and the DPS singularity vanishes. It was shown that the prefactors of $\log(\vect{Q_2}^2)$ in the diagrams with
DPS divergences could be rationalised as the products of square rooted helicity dependent splitting functions
(the prefactors of the scalar gluon and 6D $\phi^3$ boxes could also be understood in this way).

We also applied our general rules to explain why the particular MHV and NMHV six-photon amplitudes discussed in 
\cite{Bern:2008ef,Bernicot:2008nd} contain no DPS divergence. The MHV amplitude does not contain a DPS divergence
since its hard matrix elements correspond to diagrams in which a $q\bar{q}$ pair go to two photons of the same
helicity, which are zero according to the MHV rules for QED. There is no DPS divergence in the NMHV amplitude
because the total $J_z$ of the initial state is not zero. We pointed out that no NMHV six-photon diagram can
ever contain a DPS divergence, whilst there are MHV helicity amplitudes that do contain a DPS divergence.

Finally, we showed that the behaviour of the DPS singular part of a one-loop amplitude (i.e. the close-to-collinear 
part) is inconsistent with the behaviour anticipated by the `dPDF framework' of Snigirev \cite{Snigirev:2003cq} for 
calculating proton-proton DPS. On the other hand, it is entirely consistent with the `2pGPD framework'
of Diehl and Schafer \cite{Diehl:2011tt}. The root of the problem in the dPDF framework is the assumption that a 2pGPD may 
be approximately factorised into a longitudinal and a transverse piece, which is not valid. The dPDF 
objects of \cite{Zinovev:1982be} are the integrals of the $\vect{b}$-space 2pGPDs over 
$\vect{b}$. These are not directly accessed in proton-proton DPS -- however, they seem to be directly
accessed in proton-heavy nucleus DPS.

\section*{Acknowledgements}

JG acknowledges financial support from the UK Science and Technology Facilities Council. All the figures
in this paper were generated using \vb{JaxoDraw} \cite{Binosi200476}.

\bibliography{boxint}{}
\bibliographystyle{JHEP}

\end{document}